# Improvement of the Standard Characterization Method on $k_{33}$ Mode Piezoelectric Specimens

Yoonsang Park[1,*], Yuxuan Zhang[1], Maryam Majzoubi[1], Timo Scholehwar[2], Eberhard Hennig[2], and Kenji Uchino[1]

1) International Center for Actuators and Transducers (ICAT), The Pennsylvania State University, University Park, PA, 16802, USA

2) R&D Department, PI Ceramic GmbH, Lindenstrasse, 07589 Lederhose, Germany

**Abstract:** Even though standard method to determine physical parameters of piezoelectric materials has been set up for several decades, bare attention has been made on loss determination method. Furthermore, several deficits have been recognized in the standard method for $k_{33}$ mode. In this study, detailed discussion on deficits of IEEE Standard $k_{33}$ is investigated and the method to resolve such deficits will be introduced. The standard $k_{33}$ specimen suffers from small capacitance, which causes huge experimental error, intrinsic electrical energy leakage, specimen setup issue and inability to directly determine "intensive" elastic properties. In order to resolve these issues, partial electrode method was introduced, and curve fitting was demonstrated to determine physical parameters and losses.

1. **Introduction**

Piezoelectric materials are widely commercialized in ultrasonic motors and transformers, since they can provide 1/20th the volume with equivalent power compared to conventional electromagnetic devices in micro scale (less than 1 cm$^3$) [1-4]. However, the roadblock toward further miniaturization of high power density piezoelectric devices has been identified as heat dissipation [5-7]. It is known that the heat dissipation in piezoelectric materials is originated from three types of losses: dielectric ($\tan \delta$), elastic ($\tan \phi$) and piezoelectric ($\tan \theta$) losses [6-9]. Therefore, to further advance the high-power density, it is essential to elucidate heat dissipation mechanism. Furthermore, integrating more reliable values of these three losses will increase the Finite Element Method (FEM) computer simulation accuracy, when compared to real experimental data.

Each loss factor is mainly divided into intensive and extensive subgroups. Thermodynamically, intensive parameters are those that depend on the size of the system, whereas extensive parameters are those that are independent of the size. Equivalently, in piezoelectric materials, intensive parameters, such as stress ($X$) and electric field ($E$), refer to externally controllable parameters, while extensive parameters, such as strain ($x$) and dielectric displacement ($D$), are the ones that intrinsically determined by the material itself. The intensive and extensive elastic, dielectric, and piezoelectric losses are defined as imaginary part of the complex parameters in the following constitutive equations in the 1D expressions, and can be written as [6]:

**Intensive constitutive equation**

$$\begin{pmatrix} x \\ D \end{pmatrix} = \begin{pmatrix} s^E & d \\ d & \varepsilon_0 \varepsilon^X \end{pmatrix} \begin{pmatrix} X \\ E \end{pmatrix} \quad (1)$$

**Extensive constitutive equation**

$$\begin{pmatrix} X \\ E \end{pmatrix} = \begin{pmatrix} c^D & -h \\ -h & (\frac{1}{\varepsilon_0})\kappa^x \end{pmatrix} \begin{pmatrix} x \\ D \end{pmatrix} \quad (2)$$

$$\varepsilon^{X*} = \varepsilon^X(1 - j\tan\delta') \quad (3)$$

$$s^{E*} = s^E(1 - j\tan\phi') \quad (4)$$

$$d^* = d(1 - j\tan\theta') \quad (5)$$

$$\kappa^{x*} = \kappa^x(1 + j\tan\delta) \quad (6)$$

$$c^{D*} = c^D(1 + j\tan\phi) \quad (7)$$

$$h^* = h(1 + j\tan\theta) \quad (8)$$

where $\varepsilon^X$ is relative dielectric permittivity in constant stress condition, $\kappa^x$ is relative inverse dielectric permittivity in constant strain condition, $s^E$ is elastic compliance in constant electric field condition, $c^D$ is stiffness in constant dielectric displacement condition, $d$ is piezoelectric constant, and $h$ is inverse piezoelectric constant. Here, superscript star ($*$) means complex notation and $j$ is imaginary notation. The intensive losses (prime) should have negative sign, so that it may maintain the positive sign when experimentally determined. The extensive losses may also maintain positive sign, considering normal response of extensive parameter (dielectric displacement, strain) to the intensive parameter (electric field, stress).

The method to determine physical parameters in various vibration modes of piezoelectric materials is already listed in IRE Standard on Piezoelectric Crystals [10] and IEEE Standard on Piezoelectricity [11]. However, there are numerous issues in this standard method. The main problem of the standard method is that it considers only one elastic loss factor, stating that the mechanical quality factor at resonance frequency ($Q_A$) is equal to that at antiresonance frequency ($Q_B$) [11]. This statement has been proved to be erroneous, in that many experimental admittance/impedance curves from various literature showed discrepancies between $Q_A$ and $Q_B$, especially that $Q_B$ is much larger than $Q_A$ in lead zirconate titanate (PZT) [12]. It was shown in the recent study that this discrepancies are from "piezoelectric loss", which the current IEEE Standard does not consider [12]. More specifically, there are serious deficits in IEEE Standard on the $k_{33}$ mode piezoelectric bar, such as small capacitance that causes huge experimental error via small current measurement [13-14] and fringing electric field effect that intrinsically causes overestimation or underestimation of related physical parameters [15-16,22-23]. Also, the specimen setup method, with attaching or not attaching the wires to standard $k_{33}$ mode, has been an issue [17]. Furthermore, with the IEEE Standard's electrical excitation method, only direct determination of extensive elastic compliance $s_{33}^D$ and its corresponding elastic loss is possible [6,18]: intensive elastic compliance and loss can be indirectly determined through the calculation, which causes error accumulation due to error propagation. The disadvantage of error propagation was rather evidently shown in the previous study, where we inevitably reported rather high errors $\pm$ 100 % on piezoelectric loss ($\tan \theta'_{33}$) in our previous paper [19].

In this paper, we specify the issues of IEEE Standard on $k_{33}$ mode piezoelectric plate and provide new and improved method to determine $k_{33}$ mode-related physical parameters and losses, by using different electrode configuration – so called "partial electrode (PE)" configuration.

## 2. Material and Methods

*2.1 Finite Element Analysis*

In order to investigate fringing field effect on Standard $k_{33}$ specimen, numerical approach was made by using a Finite Element Analysis (FEA) software (ATILA++, distributed by Micromechatronics Inc., State College, PA). In order to see the effect of fringing field, we compared two cases: piezoelectric material with and without air background, which has relative permittivity of 1 (Refer to Table 1. for parameters used in the simulations). For the sample geometry, square dimension of top and bottom electrode was utilized, so that the width and the thickness are the same. The length of the geometry was kept as 20 mm and the width and the thickness were varied, so that the length to width aspect ratio can be set as variable parameter. For each simulation, the mesh size was kept as 50 × 4 × 4, much larger mesh numbers in the length, which is the polarization and mechanical wave propagation direction. For simulation including background, the background volume of 60 mm × (width + 40) × (thickness + 40) was set for each different geometry. The mesh for the background was set to 58 × 12 × 12.

*2.2 Experiment*

5 types of samples of PIC 255 [PI Ceramic GmbH, Lederhose, Germany], soft PZT, with the dimension of 20 × 2.5 × 0.5 mm, were prepared: IEEE Standard $k_{31}$ specimen, IEEE Standard $k_{33}$ specimen, PE open circuit (OC), PE short circuit (SC), and PE side electrode (SE). For geometry of samples, refer to Figure 1 for standard samples and to Figure 5. For PE samples. For all the samples, Au was used for the electrodes. For PE samples, center electrode was maintained to 10 % of the total length of the sample (width variation ± 20 %). The PE samples were prepared by poling a bulk ceramic block, cutting it into pieces with the specifies dimension, creating center electrode and re-poling the center part. The off-resonance and on-resonance admittance/impedance spectra for each sample were measured with 4294A Precision

Impedance Analyzer [Agilent Technologies, Santa Clara, CA]. For standard $k_{33}$ samples, in order to minimize the electrical noise, constant peak voltage of 1 V, which is the maximum voltage for the usual impedance analyzer, was applied to the specimens; no peak distortion nor heat generation (less than 0.2°C) was observed though maximum voltage was applied [5]. Also, for Standard $k_{33}$ samples, the wires were attached to each edge of the sample and the sample was measured; without wires, the resonance peaks were severely damped and could not be measured reliably. For all other PE samples and standard $k_{31}$ samples, constant peak voltage of 100 mV was applied, in order to avoid distortion of resonance peaks.

## 3. Theory

*3.1 Determination of Real Physical Parameters*

The real physical parameter determination method for Standard $k_{33}$ mode piezoelectric plate is well established in IEEE Standard on Piezoelectricity [11]. Suppose that $k_{33}$ mode piezoelectric plate, as shown in Figure 1 (a), that has the dimension of length ($l$), width ($w$), and thickness ($t$). Far from resonance regime, piezoelectric bar is treated as being purely capacitive. Therefore, stress-free relative dielectric permittivity ($\varepsilon_{33}^X$) can be determined from off-resonance impedance measurement, typically at 1 kHz, using the following equations:

$$C = \frac{1}{\omega |Z|} \quad (9)$$

$$\varepsilon_{33}^X = \frac{Cl}{\varepsilon_0 wt} \quad (10)$$

Where $C$ is capacitance, $\omega$ is angular frequency ($\omega = 2\pi f$, $f =$ frequency), $Z$ is impedance, and $\varepsilon_0$ is free-space permittivity. Because of intrinsically small capacitance of standard $k_{33}$ mode piezoelectric specimen, $\varepsilon_{33}^X$ is often measured by using standard $k_{31}$ piezoelectric specimen, which is shown in Figure 1 (b). From on-resonance admittance spectrum, extensive elastic compliance ($s_{33}^D$) and electromechanical coupling factor ($k_{33}$) can be determined, using the following equations:

$$s_{33}^D = \frac{1}{4\rho f_B^2 l^2} \tag{11}$$

$$k_{33}^2 = \frac{\pi}{2} \frac{f_A}{f_B} \tan\left(\frac{f_B - f_A}{f_B}\right) \tag{12}$$

where $\rho$ is mass density, $f_A$ is resonance frequency and $f_B$ is antiresonance frequency. These frequencies are maximum admittance and maximum impedance point, respectively. After determining $s_{33}^D$ and $k_{33}$, intensive elastic compliance ($s_{33}^E$), extensive dielectric permittivity ($\varepsilon_{33}^x$), and piezoelectric constant ($d_{33}$) can be determined sequentially:

$$s_{33}^E = \frac{s_{33}^D}{(1 - k_{33}^2)} \tag{13}$$

$$\varepsilon_{33}^x = \varepsilon_{33}^X (1 - k_{33}^2) \tag{14}$$

$$d_{33}^2 = k_{33}^2 (\varepsilon_0 \varepsilon_{33}^X s_{33}^E) \tag{15}$$

*3.2 Determination of Imaginary Physical Parameters (losses)*

Modifying the Mezheritsky's treatment [20], Uchino *et al*. [4] proposed a user-friendly methodology for determination of loss factors (imaginary physical parameters) of typical piezoelectric ceramic with *6mm* point group symmetry. First, the intensive dielectric loss factor ($\tan \delta'_{33}$) can be directly determined from the phase lag measurement at an off-resonance low frequency (such as 1 kHz). In Standard $k_{33}$ specimen, the corresponding elastic loss from $Q_B$ for $s_{33}^D$ is not exactly $\tan\phi_{33}$ (though practically close) due to mechanical constraint [19], but it can be represented as triple prime loss, which is:

$$\tan\phi_{33}''' = \frac{1}{Q_{B,33}} = \frac{1}{1-k_{33}^2}[\tan\phi_{33}' - k_{33}^2(2\tan\theta_{33}' - \tan\delta_{33}')] \tag{16}$$

where $Q_{B,33}$ is mechanical quality factor of $k_{33}$ mode at antiresonance frequency. $\tan\phi_{33}$ can rather be directly determined from $Q_B$ of thickness mode ($k_t$) sample. Mechanical quality factors can be determined with half-power bandwidth by using either 3 dB method or quadrantal bandwidth method [6,9]. After additionally measuring mechanical quality factor at resonance frequency ($Q_{A,33}$), intensive elastic loss ($\tan\phi_{33}'$) and intensive piezoelectric loss ($\tan\theta_{33}'$) can be determined by solving the following two equations [19].

$$\tan\phi_{33}' - 2k_{33}^2\tan\theta_{33}' = \frac{(1-k_{33}^2)}{Q_{B,33}} - k_{33}^2\tan\delta_{33}' \tag{17}$$

$$-\tan\phi_{33}' - 2\tan\theta_{33}' \tag{18}$$

$$= \left(\frac{1}{Q_{A,33}} - \frac{1}{Q_{B,33}}\right)\frac{(k_{33}^2 - 1 + \Omega_{A,33}^2/k_{33}^2)}{2} + \tan\delta_{33}'$$

where $\Omega$ is normalized frequency, defined as:

$$\Omega = \frac{\omega l}{2v_{33}^D} \tag{19}$$

$$v_{33}^D = \frac{1}{\sqrt{\rho s_{33}^D}} \tag{20}$$

and $v_{33}^D$ is sound velocity determined from extensive elastic compliance. Therefore, $\Omega_{A,33}$ and $\Omega_{B,33}$ are normalized resonance and antiresonance frequency, respectively. The other remaining parameter, such as extensive dielectric loss ($\tan\delta_{33}$) and extensive piezoelectric loss ($\tan \theta_{33}$), can be determined with the aid of $k_t$ mode sample.

4. Results and Discussion

*4.1 Issues of IEEE Standard $k_{33}$ Specimen*

*4.1.1 Small Capacitance*

According to IEEE Standard on Piezoelectricity for $k_{33}$ mode, the length to thickness or width ratio (*l*/*w* or *l*/*t*) should be sufficiently large that the boundary condition in minor surface can be ignored [10-11]. This is because piezoelectric bar cannot fully show the properties of $k_{33}$ mode due to mode coupling when *l*/*w* or *l*/*t* is not large enough [21]. From this statement, there exist a dilemma: in order to eliminate mode coupling, *l*/*w* or *l*/*t* should be large enough, but, as *l*/*w* or *l*/*t* increases, the capacitance of the bar becomes smaller that it can cause enormous experimental error and electrical noise. Figure 2 shows electrical response under the constant voltage drive of PIC 255 IEEE Standard $k_{33}$ mode specimen, with the dimension of 20 × 2.5 × 0.5 mm. For Figure 2, the peak voltage of 1 V, typical impedance analyzer's maximum voltage, was applied in order to minimize the electrical current noise. As shown in Figure 2 (a), there is no electrical noise and admittance curve shows a clear shape, due to high admittance value near $f_A$. However, as shown in Figure 2 (b), the impedance curve shows huge current noise because of high impedance near $f_B$. Figure 2 (c) shows dielectric permittivity in off-resonance range (800 – 1200 Hz). Even in off-resonance regime, due to small capacitance, the permittivity plot also shows huge electrical noise.

These huge electrical noises caused by intrinsic structure of Standard $k_{33}$

specimen drastically affect the parameter values and experimental errors. These noises not only prevent exact determination of parameters, but also significantly increase the error range. Exact parameter determination is very challenging for $\varepsilon_{33}^X$ due to significant electrical noise in off-resonance regime. This issue may partially be solved, if $\varepsilon_{33}^X$ is measured with a $k_{31}$ mode specimen (See Figure 1 (b)) of the same material that has much larger capacitance. However, as $s_{33}^D$ and $\tan\phi_{33}'''$ are determined from impedance curve near half-power bandwidth (3 dB point) of $f_B$, there would be huge experimental error in those parameters. Especially, error for $\tan\phi_{33}'''$ becomes enormously large, since the percentage error for the gap between 3 dB points is largely affected by the noise, due to its much smaller value compared to $f_B$ itself. It is expected that this problem becomes much more critical in hard type piezoelectric materials, because they usually have much narrower 3 dB gap than soft piezoelectric materials do (thus much higher $Q_{B,33}$), considering similar noise level. Furthermore, more errors will accumulate for the other parameters, especially intensive loss factors, due to error propagation.

*4.1.2 Fringing Electric Field*

As already mentioned in the previous section, the standard $k_{33}$ specimen inevitably has large ratio of *l/w* or *l/t*. However, there is another significant issue caused by this large aspect ratio: fringing electric field. As already dealt in basic electrostatic problems, if the separation between two parallel conducting plates becomes much larger than the dimensions of the plates itself, the electric field leaks and parameter values can be distorted, such as overestimation of the dielectric permittivity [15-16,22-23]. However, the effect of fringing field becomes much less, if a material's dielectric permittivity is much larger than the background. This is because the electric field must be effectively ducted through high permittivity material.

Normally, piezoelectric materials have much higher permittivity ($\varepsilon_r$ = 1000 ~ 2000) compared to air ($\varepsilon_r$ = 1); therefore, the effect of fringing field is likely to be small. However, in the standard $k_{33}$ specimen, there should remain fringing field effect, since *l/w* or *l/t* is very large. In order to investigate how *l/w* or *l/t* affects the magnitude of fringing field and distortion of parameters, numerical approach was made by using ATILA FEA software. Table 1 shows physical properties of both piezoelectric materials and air background. For piezoelectric material, we utilized physical parameters of PZT 5A (Soft PZT), which was already built in the ATILA software.

Figure 3 shows how much electric field leaks and how much can $\varepsilon_{33}^X$, $s_{33}^D$, and $k_{33}$ be affected by increasing *l/w* of standard $k_{33}$ specimen. The inset in each plot magnifies each corresponding plot when the *l/w* value ranges from 3 to 7, because these aspect ratios are widely utilized for standard $k_{33}$ specimen [17,24-25]. All the plots in Figure 3 exclude the values at aspect ratio of 1 and 2, since these aspect ratios does not show pure $k_{33}$ mode due to mode coupling. For Figure 3 (b), (c) and (d), the FEA calculation results from geometry with and without air background was compared in order to distinctly show the effect of fringing field on these parameters. Figure 3 (a) shows the percentage of electric energy leak, which is defined by $c_f/(c_f + c_p)$, where $c_f$ is floating capacitance due to fringing effect and $c_p$ is capacitance of parallel conducting plates. The leakage percentage nonlinearly increases, as *l/w* increases. Figure 3 (b) shows how $\varepsilon_{33}^X$, changes by increasing *l/w*. Without background, $\varepsilon_{33}^X$ is fixed to 1700, since FEA calculates parallel plate capacitance when background is not included. However, when background is included in the calculation, it is shown that $\varepsilon_{33}^X$ is overestimated, and the level of overestimation drastically increases as *l/w* increases. This behavior was already explained by many studies [15-16,26-27]. Figure 3 (c) and (d) shows *l/w* dependence of $s_{33}^D$ and $k_{33}$ with and without air background. For both parameters when background is not included, slight upturn at *l/w* of 3 occurs, because of the slight mechanical clamp from the width/thickness direction; however, the clamp effect becomes smaller as *l/w* increases higher than 7, and both parameters merges into constant value when *l/w* becomes adequately large. From those figures, it is noted that $k_{33}$ is underestimated and $s_{33}^D$ is overestimated, when electric energy leakage is considered. This is because the $f_B$ shifts toward lower frequency (due to floating capacitance) while $f_A$ is maintained constant, when *l/w* increases gradually.

It is noteworthy that the behaviors of $\varepsilon_{33}^X$, $s_{33}^D$, and $k_{33}$ with respect to *l/w* correspond to the electric energy leakage percentage, as shown in Figure 3. The drastic increase of overestimation of $\varepsilon_{33}^X$, $s_{33}^D$ and drastic underestimation of $k_{33}$ are in good agreements with drastic increase of leakage energy with gradual increase of *l/w*. The aspect ratio range shown in each inset represents the mostly adopted aspect ratio for standard $k_{33}$ specimen. Even though the intrinsic errors caused by fringing field effect ranges from about 0.5 percent (*l/w* = 3) to about 2 percent (*l/w* = 7), more error will be accumulated when other parameters are calculated. It is recommended that these values should be calibrated according to the aspect ratio.

*4.1.3 Specimen Setup Issue*

Standard $k_{33}$ specimen not only suffers from intrinsic errors, such as small capacitance and electric energy leakage, but also possesses experimental challenge. When the standard $k_{33}$ specimen is measured with commercially available impedance analyzer, the specimen is sandwiched between two metallic sample holder tips of fixture. However, this method is problematic especially for standard $k_{33}$ specimen, since the tips sandwiching the sample partially clamp or apply force in the vibration direction, damping the vibration of the specimen. As a result, resonance and antiresonance peaks are severely damped, and exact parameter determination remains challenging.

Attaching wires to standard $k_{33}$ specimen can partially resolve this issue, because the specimen is not sandwiched by the fixture tips anymore. However, attaching wires to the sample means adding mass and rigidity; therefore, it affects the parameter determination. The issue caused by the mass of the wires may be solved by measuring the specimen with different wire lengths and extrapolate the parameter values. This method was motivated by Slabki *et al*. [17], which compared the parameter values of standard $k_{33}$ specimen in two cases: with and without wires. The measurements were taken by originally attaching two wires of 10 cm to the specimen on each side and cutting the wires by 2 cm increment. After parameters were determined for each wire length, the data was fitted with weighted linear regression line and each parameter was extrapolated to zero wire length.

Figure 4 summarizes measured values of $s_{33}^D$ and $k_{33}$ with different wire lengths, showing that $s_{33}^D$ is overestimated and $k_{33}$ is underestimated, when wires are attached to the specimen. The extrapolated values from linear fitting for $s_{33}^D$ and $k_{33}$ were determined to be 8.88 μm²/N and 0.717, respectively. The overestimation of $s_{33}^D$ and underestimation of $k_{33}$ are due to that the $f_B$ is shifted toward lower frequency when the wires are attached. No changes in $f_A$ were observed with wire attachment. This is because $f_B$ represents the mechanical resonance for standard $k_{33}$ mode [6,18]. For $k_{33}$ mode, the given electrical boundary condition is "*D* - constant"; therefore, extensive elastic compliance ($s_{33}^D$), which is determined from $f_B$, should be considered, rather than intensive elastic compliance ($s_{33}^E$). Even though the parameter distortion caused by wire attachment can be solved by this wire extrapolation method, more problematic is the loss determination. Figure 4 (c) and 4 (d) show wire length-dependence of the mechanical quality factors $Q_A$ and $Q_B$. Both $Q_A$ and $Q_B$ seem to keep constant value as wire length changes. The error in $Q_A$ is small, since Standard $k_{33}$ specimen shows almost noiseless admittance curve near $f_A$, as shown in Figure 2 (a). However, due to huge noise near $f_B$, as shown in Figure 2 (b), the error for $Q_B$ is enormously large. Because of the large fluctuation of admittance spectrum near $f_B$, we could not conclude the reliable $Q_B$ value and its behavior with respect to wire length.

*4.1.4 Inability to Directly Determine Intensive Elastic Parameters*

One of the problems of IEEE standard samples is that either extensive or intensive elastic compliance and loss can be directly determined. This is because each standard sample is bounded by one specific electrical boundary condition. For example, with standard $k_{31}$ sample, as shown in Figure 1 (b), only intensive elastic compliance ($s_{11}^E$) and loss ($\tan \phi'_{11}$) can be determined [9]. extensive elastic compliance ($s_{11}^D$) and loss ($\tan \phi_{11}$) should be indirectly determined from the calculation via so-called K-matrix. For standard $k_{33}$ sample, as shown in Figure 1 (a), the situation is opposite: extensive elastic compliance ($s_{33}^D$) and corresponding loss ($\tan \phi'''_{33}$) can be directly determined [19], while intensive elastic compliance ($s_{33}^E$) and loss ($\tan \phi'_{33}$) should be indirectly determined from calculation. The problem of limited electrical boundary condition of standard samples is that indirectly calculated compliance and loss experience huge error due to error propagation.

*4.2 Partial Electrode Configuration as a Methodology for Accurate Parameter Determination*

The PE configuration method was previously proposed by Majzoubi *et al*. [9], in order to directly determine both intensive and extensive elastic compliances ($s_{11}^E$, $s_{11}^D$) and losses ($\tan \phi'_{11}$, $\tan \phi_{11}$) for $k_{31}$ mode. The PE configuration has advantage in its mechanical excitation over the IEEE standard characterization method, in which the specimen is electrically driven. Figure 5 shows partial electrode configuration for $k_{33}$ mode, along with IEEE standard $k_{33}$ specimen (Figure 5 (a)). Each PE configuration shown in Figure 5 (b), (c) and (d) is composed of a small center electrode part, which are $k_{31}$ mode, which is used for both mechanical excitation and vibration monitoring, and the side load parts, which are $k_{33}$ mode. When the center part is electrically excited, the piezoelectric stress generated by the center part will mechanically excite the side part; therefore, by maintaining the center part ($k_{31}$ mode) smaller (about 10 %) than the side part ($k_{33}$ mode), the admittance/impedance curve reflects the properties of the side $k_{33}$ mode, and $k_{33}$ mode-related parameters can be determined with the admittance/impedance measurement at the center electrode under proper analysis. PE for $k_{33}$ mode has additional advantage: the capacitance of center part is much larger than that of standard $k_{33}$ specimen. As shown in Figure 5. (b) and (c), open circuit (OC) and short circuit (SC) configurations were designed; these configurations correspond to antiresonance and resonance drive, respectively. Furthermore, by putting electrode on the side, intensive elastic compliance and loss can be directly determined with side electrode (SE) samples; this is what cannot be conducted with IEEE Standard $k_{33}$ specimen and will significantly reduce the error for them.

More specifically, the advantage of partial electrode, though its complex structure, is rather clear. Firstly, the capacitance value is dramatically increased. Considering sample dimension of 20 × 2.5 × 0.5 mm for both standard $k_{33}$ specimen and PE, PE simply has 160 times larger static capacitance than the standard specimen does even when the center portion is about 10 % of the entire length. Therefore, admittance/impedance measurement becomes much more reliable, with significantly reduced electrical noise. Secondly, for PE, the fixture with two tips can be simply used for the measurement, since the tip does not clamp the sample in the vibration direction. The researchers do not need to suffer from multiple measurements with different wire lengths, Thirdly, the leakage effect is smaller in PE than in standard specimen, considering the side length of PE. Fourth, with side electrode configuration, intensive elastic compliance and loss can be directly determined, so that the error can be significantly reduced.

*4.3 Methodology of Parameter Determination of $k_{33}$ Mode Using Partial Electrode*

In order to determine physical parameters of $k_{33}$ mode, samples made of PIC 255, which is soft PZT with morphotropic phase boundary (MPB) composition, were used. For consistency, open circuit samples were measured first, then the wires were attached to the same samples to create short circuit samples. For short circuit sample, the wires should be attached necessarily. However, no changes in admittance/impedance curve were observed as the wire length changes. This is because short circuit configuration corresponds to resonance drive of standard sample. As already mentioned in *4.1.3*, if we attached wires to standard sample, only $f_B$ was shifted and $f_A$ remained the same. Therefore, no wire extrapolation was required for short circuit samples.

By considering mechanical and electrical boundary conditions for each configuration, analytical solutions for all $k_{33}$ PE configurations were derived. The detailed derivation process, along with validation of analytical solutions are described in a separate paper [28]. The admittance equations for open circuit, short circuit, and side electrode are:

$$Y_{OC} = j\omega \left[ \frac{2d_{31}^{*2} v_{11}^{E*} v_{33}^{D*} s_{33}^{D*} \cos\left(\frac{\omega}{v_{33}^{D*}} \frac{(a-1)l}{2}\right) \sin\left(\frac{\omega}{v_{11}^{E*}} \frac{al}{2}\right)}{t s_{11}^{E*} v_{33}^{E*} s_{33}^{E*} \cos\left(\frac{\omega}{v_{33}^{D*}} \frac{(a-1)l}{2}\right) \cos\left(\frac{\omega}{v_{11}^{E*}} \frac{al}{2}\right) + v_{11}^{E*} s_{11}^{E*} \sin\left(\frac{\omega}{v_{11}^{E*}} \frac{al}{2}\right) \sin\left(\frac{\omega}{v_{33}^{D*}} \frac{(a-1)l}{2}\right)} + \frac{al\omega \varepsilon_0 \varepsilon_{33}^{X*}(1 - k_{31}^{*2})}{t} \right] \quad (21)$$

$$Y_{SC} = j\omega \left[ \frac{2d_{31}^{*2} v_{11}^{E*} v_{33}^{D*} s_{33}^{D*}}{s_{11}^{E*} t \left( \frac{s_{33}^{D*} v_{33}^{D*}}{\tan\left(\frac{al\omega}{2v_{11}^{E*}}\right)} + v_{11}^{E*} s_{11}^{E*} \left\{ \frac{8 d_{33}^{*2} v_{33}^{D*} \sin^2\left(\frac{(1-a)l\omega}{4v_{33}^{D*}}\right) + l\omega(1-a)\varepsilon_0 \varepsilon_{33}^{X*} s_{33}^{E*} \sin\left(\frac{(1-a)l\omega}{2v_{33}^{D*}}\right)}{2 d_{33}^{*2} v_{33}^{D*} \sin\left(\frac{(1-a)l\omega}{2v_{33}^{D*}}\right) - l\omega(1-a)\varepsilon_0 \varepsilon_{33}^{X*} s_{33}^{E*} \cos\left(\frac{(1-a)l\omega}{2v_{33}^{D*}}\right)} \right\} \right)} + \frac{al\omega \varepsilon_0 \varepsilon_{33}^{X*}(1 - k_{31}^{*2})}{t} \right] \quad (22)$$

$$Y_{SE}$$
$$= jw \left[ \frac{2d_{31}^{*2} v_{11}^{E*} v_{33}^{E*} s_{33}^{E*} \cos\left(\frac{\omega}{v_{33}^{E*}} \frac{(a-1)l}{2}\right) \sin\left(\frac{\omega}{v_{11}^{E*}} \frac{al}{2}\right)}{ts_{11}^{E*} v_{33}^{E*} s_{33}^{E*} \cos\left(\frac{\omega}{v_{33}^{E*}} \frac{(a-1)l}{2}\right) \cos\left(\frac{\omega}{v_{11}^{E*}} \frac{al}{2}\right) + v_{11}^{E*} s_{11}^{E*} \sin\left(\frac{\omega}{v_{11}^{E*}} \frac{al}{2}\right) \sin\left(\frac{\omega}{v_{33}^{E*}} \frac{(a-1)l}{2}\right)} \right.$$
$$\left. + \frac{al\omega\varepsilon_0 \varepsilon_{33}^{X*}(1-k_{31}^{*2})}{t} \right] \quad (23)$$

Where $Y_{OC}$, $Y_{OC}$ and $Y_{OC}$ are admittance for PE open circuit, short circuit, and side electrode, respectively, $l$, $w$, $t$ are length, width and thickness of the plate, respectively, $\omega$ is angular frequency, and $a$ is portion of the center electrode, ranging from 0 to 1. Therefore, each parameter contains the real and imaginary (loss) parameter within itself.

For the parameter determination process, curve fitting method was utilized, as already shown in many cases for parameter determination [29-31]. The parameter determination process is the following: first of all, the exact dimension and mass of each sample were measured, in order to obtain "$\rho$". Also, the dimensions of the center electrode of each sample were measured in order to obtain "$a$", the percentage portion of the center electrode. This can be done with the help of optical microscope. Also, because the analytical solution contains $k_{31}$ mode related parameters, standard $k_{31}$ sample with the same material (PIC 255) were measured and related parameters were determined with the method explained in Uchino *et al*. [6]. The determined parameters of standard $k_{31}$ sample are shown in Table 2. Note that the losses are determined by the first approximation of Taylor series, so only to the second digit is valid, though experimental errors are smaller than the digit shown. The off-resonance permittivity and dielectric loss can be determined by measuring impedance and phase lag at 1 kHz. Since the center portion is $k_{31}$ mode, $\varepsilon_{33}^{X}$ and $\tan\delta_{33}'$ can be directly measured, without suffering from small capacitance of standard $k_{33}$ mode. If the $k_{31}$ mode-related values are plugged into the expression for $Y_{OC}$, the admittance equation for PE OC, the remaining parameters are only the $s_{33}^D$ and $\tan \phi_{33}'''$ that can be determined by curve fitting. Next, by applying similar method to SE configuration, $s_{33}^E$ and $\tan \phi_{33}'$ can be determined. Since SC samples were created by attaching the wires to the same OC samples that were already measured, by plugging in the corresponding values of $s_{33}^D$ and $\tan \phi_{33}'''$ for each corresponding sample and using $s_{33}^E$ and $\tan \phi_{33}'$ values determined from SE samples, $d_{33}$ and $\tan \theta_{33}'$ can be determined. It should be noted that, during the fitting process, about 10 % of underestimation of $d_{31}$, compared to determined value in Table 2, occurred. The reason may be due to the poling imperfection near the boundary between the center and the side. However, since $d_{31}$ is the parameter of center "actuator", not the parameter of key "side-load" and does not affect resonance frequencies and mechanical quality factors, the fitting can be considered valid method of parameter determination.

After the determination of $\varepsilon_{33}^X$, $\tan \delta_{33}'$, $s_{33}^D$, $\tan \phi_{33}$, $d_{33}$ and $\tan \theta_{33}'$, other remaining parameters can be indirectly determined. $k_{33}$ can be calculated from the following equation [4]:

$$k_{33}^2 = \frac{d_{33}^2}{s_{33}^E \varepsilon_0 \varepsilon_{33}^X} \tag{24}$$

$k_{33}^2$ can also be determined from $s_{33}^D$ and $s_{33}^E$ by the following relation, which is mathematically equivalent to Equation (13):

$$1 - k_{33}^2 = \frac{s_{33}^D}{s_{33}^E} \tag{25}$$

The coupling loss (imaginary part of $k_{33}$) can be calculated with the following equation [4]:

$$\tan \chi_{33} = 2 \tan \theta'_{33} - \tan \delta'_{33} - \tan \phi'_{33} \qquad (26)$$

For each configuration, admittance curve for 10 samples were fitted to analytical solutions. The percentage fitting error was considered for the error for the determined parameters; all the fitting errors were less than 1 %, which represents excellent fit. The fitting result is shown in Figure 6. Each percentage fitting error merged by error propagation in order to generate total error.

*4.4 Comparison of PE and Standard Sample*

Table 3 shows the comparison of experimentally determined parameters between standard $k_{33}$ sample and partial electrode. For standard sample, the parameter values were determined via wire extrapolation and the average percentage error from different wire lengths were considered. The extrapolated values are then further calibrated considering fringing field effect. ATILA FEA simulation results for the standard sample with the dimension of 20 × 2.5 × 0.5 mm shows that about 7.4 % overestimation in $s_{33}^D$ and 2.4 % underestimation of $k_{33}$. For error determination of standard $k_{33}$ samples, the electrical noise in off-resonance regime and near $f_B$ was considered. Since we cannot avoid slight fringing electric energy effect for PE samples, precisely speaking, the determined parameter values for PE should also be calibrated by considering the portion of the side part.

First of all, there is a huge difference in error of $\varepsilon_{33}^X$ and $\tan \delta'_{33}$. Standard $k_{33}$ samples provide huge error values, due to electrical noise at off-resonance region; especially, the error in $\tan \delta'_{33}$ is extremely large, so that it is not considered as reliable value. However, $\varepsilon_{33}^X$ and $\tan \delta'_{33}$ determined from PE samples gives much less errors with significantly reduce noise, and they are in similar level compared to those of Standard $k_{31}$ samples in Table 2. Next, the errors of $s_{33}^D$ and $\tan\phi_{33}'''$ of PE samples are less than those of standard sample. Though the electrical noise does not affect error of $s_{33}^D$ so much because the noise fluctuation is very small compared to the $f_B$ (~ 90 kHz) itself, it heavily affects the error of $\tan\phi_{33}'''$, since $\tan\phi_{33}'''$ is reciprocal of $Q_{B,33}$, and $Q_{B,33}$ is determined from half-power bandwidth (3 dB point); the bandwidth is only within the range of ~ 200 Hz, which was significantly affected by the noise. However, since $s_{33}^D$ and $\tan\phi_{33}'''$ were directly determined by fitting experimental admittance curve of OC samples with very small fitting errors, the errors are very small compared to those of standard samples. For $s_{33}^E$, $\tan\phi_{33}'$, $d_{33}$, and $\tan\theta_{33}'$, standard samples provide large errors due to error propagation, whereas PE samples provide much less errors, because these were directly determined by fitting experimental data of SC and SE samples. Furthermore, it is noteworthy to mention that $k_{33}$ values determined from two separate equations (71.7 % from Equation (24) and 69 % from Equation (25)) are in a good agreement. Even though $k_{33}$ was indirectly determined for PE samples, the error for $k_{33}$ came out to be smaller than that of standard sample, due to small error values in $s_{33}^E$, $\varepsilon_{33}^X$ and $d_{33}$. Finally and importantly, $\tan\chi_{33}$ of standard $k_{33}$ is massive (480 %), due to error propagation from intensive losses that already have huge error. Compared to standard $k_{33}$, $\tan\chi_{33}$ of PE is much smaller (12 %). This is due to that all the intensive loss factors have smaller errors in PE.

## 5. Conclusion

In this study, we pointed out serious deficits in IEEE Standard $k_{33}$ specimen and proposed PE as better method to determine parameters and loss factors of $k_{33}$ mode piezoelectric plate. Because the PE configuration utilizes mechanical excitation, the electrical energy leakage problem in IEEE Standard method can be eliminated, and both *E*- and *D*- constant conditions can be adopted on the side mechanical load $k_{33}$ specimens, leading to direct determination of both $s_{33}^D$ and $s_{33}^E$. The benefit of PE also included that it provides much higher capacitance at the center actuator/monitoring electrode part, so that intensive elastic compliance and elastic loss factor can be directly determined to give less error. Furthermore, PE method was confirmed by comparing both $k_{33}$ values determined from two different equations. The exact loss determination is essential to better understand heat dissipation mechanism in piezoelectric materials and provide more efficient way (less heat generation) to drive the high-power piezoelectric devices.

**Acknowledgement**

This work was supported by Office of Naval Research under Grant Number N00014-17-1-2088.

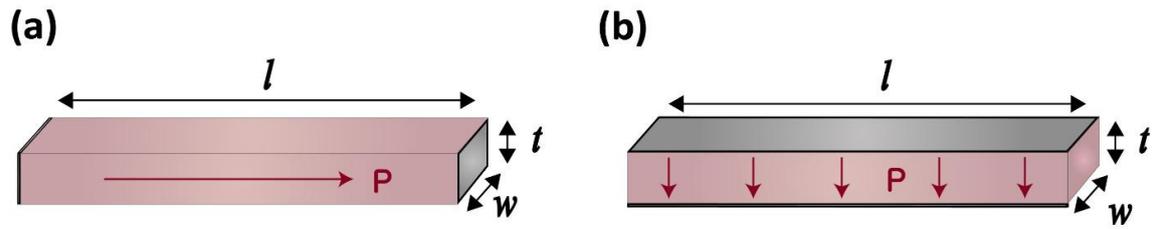

Figure 1. (a) IEEE Standard $k_{33}$ mode and (b) IEEE Standard $k_{31}$ mode piezoelectric plate

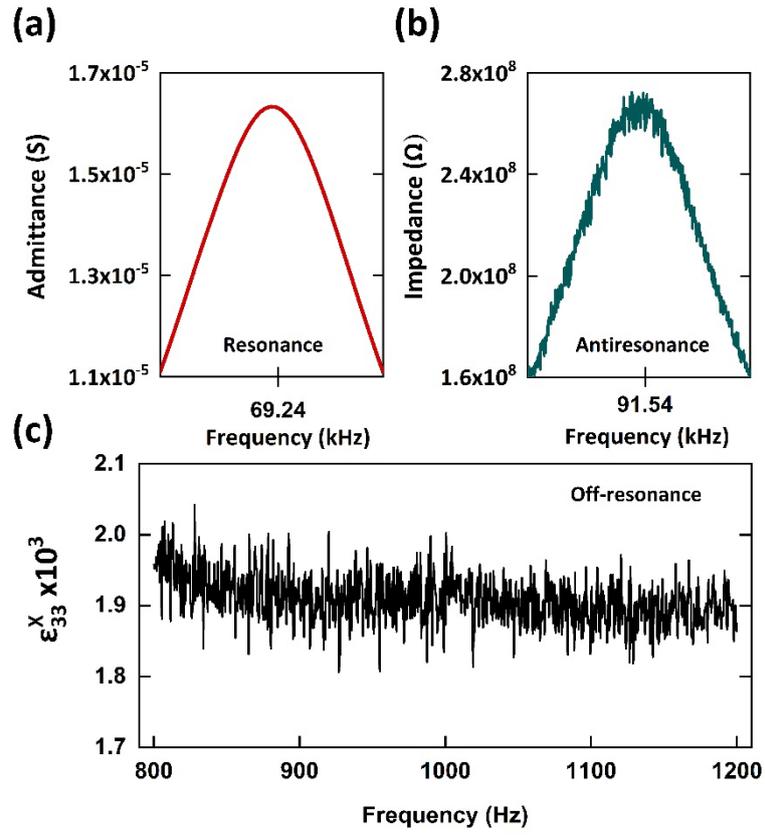

Figure 2. (a) Admittance curve of resonance frequency and (b) impedance curve of antiresonance frequency of PIC 255 standard $k_{33}$ specimen near half power bandwidth (3 dB point). (c) Relative permittivity ($\varepsilon_{33}^X$) vs. frequency plot in off-resonance region (800 – 1200 Hz).

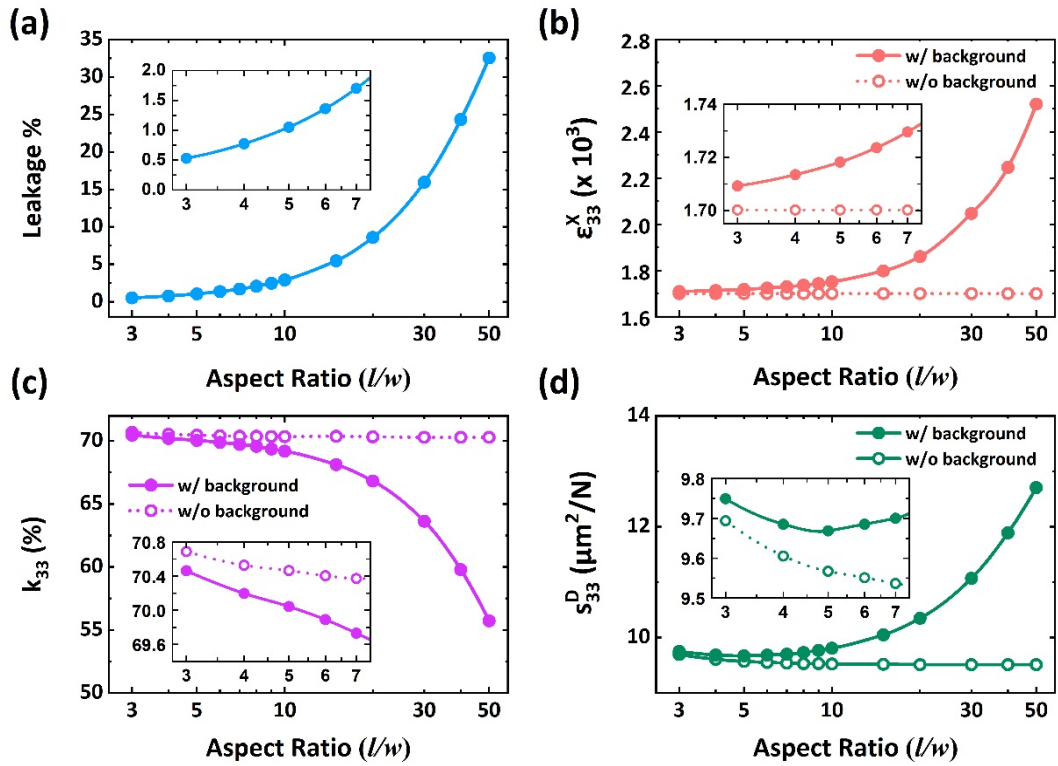

Figure 3. (a) The percentage of electric energy leak (Blue), (b) $\varepsilon_{33}^X$ (Red), (c) $k_{33}$ (Purple) and (d) $s_{33}^D$ (Green) as functions of aspect ratio (*l/w*) calculated from ATILA FEA in two distinct cases, with (Solid circle) and without air background (Open circle). Each inset in Figure 3. (b), (c) and (d) shows magnification at *l/w* of from 3 to 7, which is widely adopted for standard $k_{33}$ specimen.

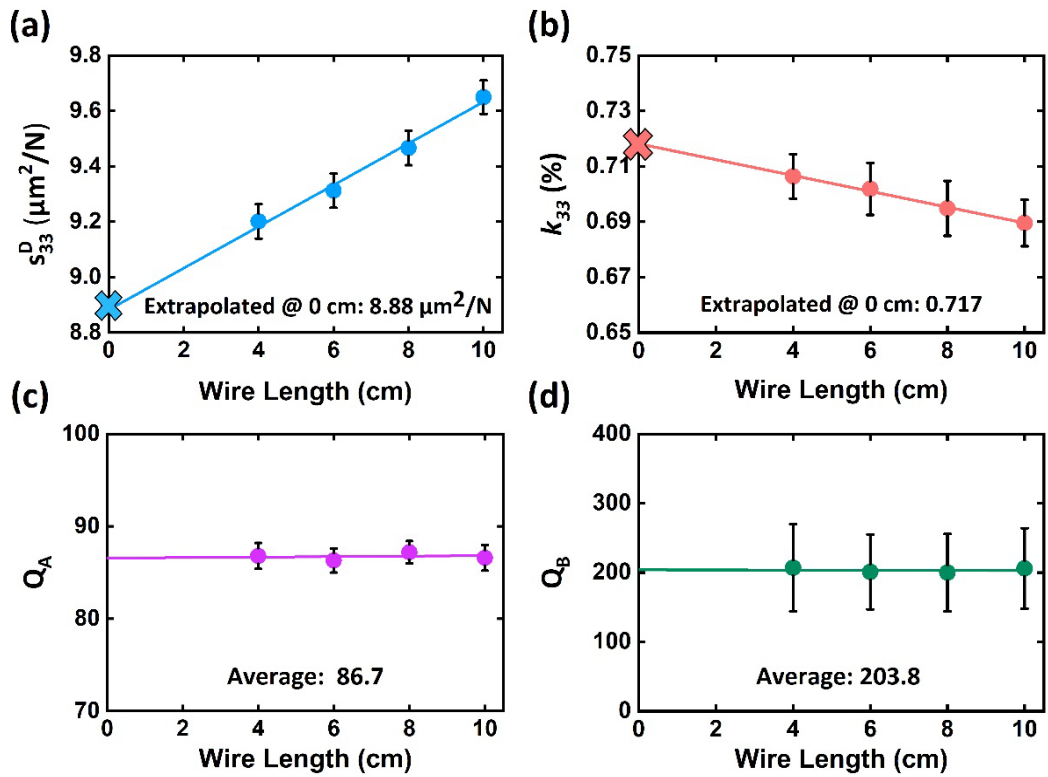

Figure 4. (a) Experimental $s_{33}^D$ (Blue solid circle), (b) $k_{33}$ (Red solid circle) (c) $Q_A$ (Purple solid circle) and (d) $Q_B$ (Green solid circle) of Standard $k_{33}$ specimen as functions of wire length. The lines in each plot shows linear regression fitting. For Figure 4 (a) and (b), the values at zero wire length were determined from regression line, to eliminate the effect of attached wires. X-mark represents zero wire extrapolation. For Figure 4 (c) and (d), the average value was considered due to no wire length-dependence of parameters.

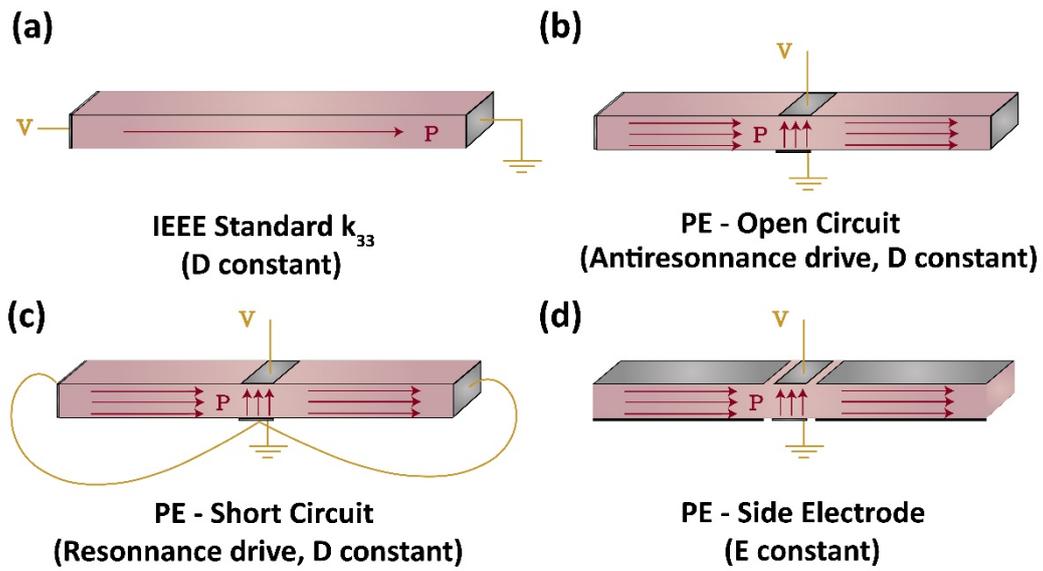

Figure 5. (a) IEEE standard $k_{33}$ specimen, (b) PE open circuit for antiresonance drive, (c) PE short circuit for resonance drive, and (d) PE side electrode for direct determination of intensive elastic compliance and loss.

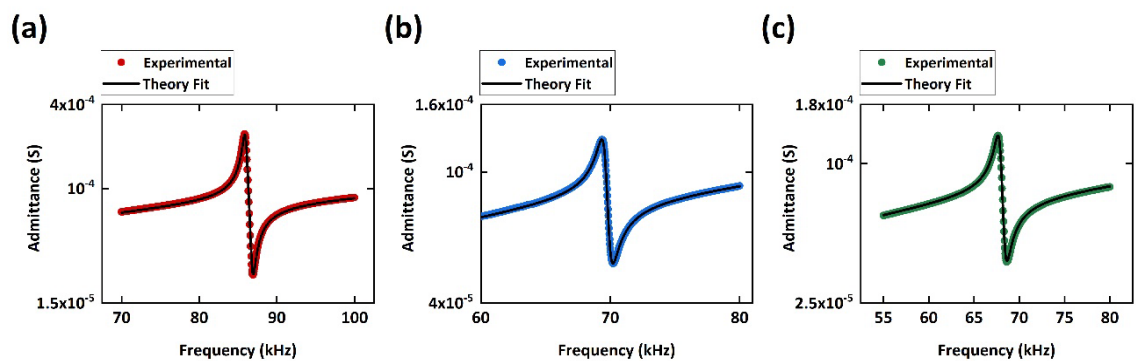

Figure 6. Experimental admittance curves for the $k_{33}$ mode partial electrodes: (a) OC (red dots), (b) SC (blue dots) and (c) SE (green dots) samples. Black lines denote analytical theory fit to each admittance curve.

|  | Piezoelectric (PZT 5A) | Background (Air) |
|---|---|---|
| Density (kg/m³) | 7780 | 1.225 |
| Relative Permittivity | $\varepsilon_{33}^X = 1700$ | 1 |
| Elastic Compliance (μm²/N) | $s_{33}^E = 18.8$ | 1 |
| Piezoelectric Constant (pC/N) | $d_{33} = 374$ | 0 |

Table 1. Physical properties of piezoelectric material (PZT 5A) and air background used in ATILA FEA simulation for Standard $k_{33}$ specimen.

## Parameters for Standard $k_{31}$ sample

| Real Parameter | | | |
|---|---|---|---|
| $\varepsilon_{33}^X$ | $s_{11}^E$ | $d_{31}$ | $k_{31}$ |
| …. | (μm²/N) | (pC/N) | (%) |
| 1808 ± 0.1 % | 16.1 ± 0.1 % | -187 ± 4 % | 0.37 ± 4 % |

| Imaginary Parameter | | | |
|---|---|---|---|
| $\tan \delta_{33}'$ | $\tan \phi_{11}'$ | $\tan \theta_{31}'$ | $\tan \chi_{31}$ |
| (%) | (%) | (%) | (%) |
| 1.6 ± 1 % | 1.2 ± 2 % | 2.2 ± 6 % | 1.6 ± 16 % |

Table 2. Physical parameters determined from standard $k_{31}$ sample that were used for curve fitting of PE samples.

**Parameters Determined from PE Sample**

| Real Parameter | | | | | |
|---|---|---|---|---|---|
| $\varepsilon_{33}^X$ …. | $s_{33}^D$ (μm²/N) | $s_{33}^E$ (μm²/N) | $d_{33}$ (pC/N) | $k_{33}$ from Eq. (24) (%) | $k_{33}$ from Eq. (25) (%) |
| 1836 ± 0.2 % | 9.21 ± 0.3 % | 17.7 ± 0.3 % | 382 ± 0.3 % | 71.7 ± 0.2 % | 69 ± 0.4 % |

| Imaginary Parameter | | | | |
|---|---|---|---|---|
| $\tan \delta_{33}'$ (%) | $\tan \phi_{33}'''$ (%) | $\tan \phi_{33}'$ (%) | $\tan \theta_{33}'$ (%) | $\tan \chi_{33}$ (%) |
| 1.7 ± 4.8 % | 0.56 ± 0.6 % | 1.2 ± 0.3 % | 1.8 ± 0.6 % | 0.7 ± 12 % |

**Parameters Determined from Standard $k_{33}$ Samples**

| Real Parameter | | | | |
|---|---|---|---|---|
| $\varepsilon_{33}^X$ …. | $s_{33}^D$ (μm²/N) | $s_{33}^E$ (μm²/N) | $d_{33}$ (pC/N) | $k_{33}$ (%) |
| 1750 ± 7 % | 8.3 ± 0.7 % | 17.8 ± 2.2 % | 389 ± 2.6 % | 73 ± 1.4 % |

| Imaginary Parameter | | | | |
|---|---|---|---|---|
| $\tan \delta_{33}'$ (%) | $\tan \phi_{33}'''$ (%) | $\tan \phi_{33}'$ (%) | $\tan \theta_{33}'$ (%) | $\tan \chi_{33}$ (%) |
| 1.8 ± 83 % | 0.49 ± 29 % | 1.2 ± 50 % | 1.8 ± 67 % | 0.6 ± 480 % |

Table 3. Physical parameters and losses determined from standard $k_{33}$ samples and PE samples.